\documentclass[aps,showpacs,twocolumn,prd]{revtex4}
\usepackage{graphicx}
\usepackage{amssymb}
\usepackage{amsmath}
\newcommand{\be}{\begin{equation}}
\newcommand{\ee}{\end{equation}}
\newcommand{\ben}{\begin{eqnarray}}
\newcommand{\een}{\end{eqnarray}}
\newcommand{\bes}{\begin{subequations}}
\newcommand{\ees}{\end{subequations}}
\newcommand{\bb}{\bibitem}

\begin{document}
\title{Low-cost educational robotics applied to physics teaching in Brazil}

\author{Marcos A. M. Souza} 
\email{msouza@ifpiparnaiba.edu.br}
\affiliation{Departamento de F\'\i sica, Instituto Federal do Piau\'\i, Parna\'\i ba, Piau\'\i, 64215-000, Brazil.}

\author{Jos\'e R. R. Duarte}
\affiliation{Departamento de F\'\i sica, Instituto Federal do Piau\'\i, Teresina, Piau\'\i, 64000-040, Brazil.}

\date{\today}\date{\today}

\begin{abstract}
In this paper we propose strategies and methodologies of teaching topics in high school physics through a show of Educational 
Robotics. The Exhibition was part of a set of actions promoted by a brazilian government program of incentive for teaching 
activities (PIBID) and whose primary focus is the training of teachers, improvement of teaching in public schools, 
dissemination of science and formation of new scientists and researchers. By means of workshops, banners and prototyping 
of robotics, we are able to create a connection between the study areas and their surrounding, making learning meaningful 
and accessible for the students involved and contributing to their cognitive development.
\end{abstract}


\maketitle

\section{Introduction }

The contemporary process of teaching and learning of physics that brings about elements based on modeling concrete objects, 
with an emphasis on empirical methods, dates back to ancient Greece. These methods of scientific investigation, which still 
today serve as guidelines in the scientific community, were popularized in the sixteenth century by names like Francis Bacon, 
William Gilbert and Galileo Galilei \cite{ref1}. From this point of view, the importance of rescuing these aspects of physics 
teaching, often designated by many authors as the triad research-action-reflection, is vital \cite{ref2}. In Brazil, 
particularly, there are few public schools that have laboratories with appropriate equipment to practical classes. Therefore, 
students say that discipline becomes far in their reality, where predominates a traditional conception of teaching 
using an archaic and tedious methodology. Mostly there is an excessive routine of lectures and list of exercises that generally 
prioritize the memorization of mathematical formulas. Because of this misguided way of teaching, many students are driven to 
mechanically repeat the solutions of similar questions previously solved by the teachers and that ends up in general not 
promoting the development of practical and cognitive skills of the students.

In this paper we reinforce the positive aspect of experimental activities in the process of teaching and learning physics 
in high school. This is the story of an exhibition of educational robotics carried out in the public schools of the state 
of Piau\'\i, in Parna\'iba, Brazil.  These activities were developed as part of the actions promoted by the PIBID (Programa
Institucional de Bolsas para Inicia\c{c}\~ao a Doc\^encia, a scholarship institutional program for teaching initiation) \cite{pibid}, 
and is an initiative of the Federal Government of Brazil in partnership with the Federal Institute of Piau\'i (IFPI). 
The program is intended to stimulate, develop and improve teaching skills of undergraduate students, having as medium to 
long-term goal the enhancement of the low-performing Basic Education of local public schools evaluated by the Ministry 
of Education (MEC). Thereby, in order to make the future teachers become aware of both the actual conditions of the 
teachers' work environment and their obligation, the teaching initiation for most undergraduates takes place under the 
supervision of their own academic teacher along with the supervision of the teachers that work in the school where 
the project is implemented. 

Note that all projects about Educational Robotics developed and presented so far were brought into reality under the 
supervision and guidance of the authors of this article and were only possible due to the financial support of the Brazilian 
Fellowship Program PIBID. Some experiments were made with recycled and low cost materials, facilitating the access of high 
school students to the practice. The results obtained from these innovating activities have already resulted in two recent 
papers \cite{ref3,ref4}.
							
The present paper is organized as follows. In Sec.II we consider the use of robotics in several contexts. Then, in Sec.III, 
we discuss the methodology used to build up an easy comprehension of the concepts involved in the field of robotics as well 
as the educational motivation for the proposal and its role in the teaching of physics. Finally, the achievements  
and conclusions are presented, respectively, in Secs.IV and V.

\section{EDUCATIONAL ROBOTICS}

The robotics branch of science involves the study and development of educational and technological architectures which 
can be simple or complex systems based on logic programming that dynamically articulate through a mechanical automaton 
structure operated by means of integrated circuits and electro hydraulic controls and tires, resulting in what is popularly 
known as Robot.

Although it is a relatively new science, officially emerged in the twentieth century, the history of robotics has its origens, 
just like physics, in ancient Greece. The seek for productive efficiency and improved quality of manufactured products have 
always been one of the main reasons for the men's interest in developing the robotics research field. For instance, 
the Pneumatic Automata is a remarkable piece of work of a Greek engineer from Alexandria and is considered one of the first 
texts on the subject. 

The term `Robot' was first mentioned in 1920 by the Czech Karel Kapec in a play entitled ``Rossum's Universal Robots" and was 
used again later, in 1950, by Isaac Asimov in his famous science fiction book: I, Robot. 

Over the past years there have been great advances in the Robotics field led by the necessity of creation of a spatial program 
and the growth of the entertainment of Lego toys, and also by the Artificial Intelligence program and the development of 
bipedal robots and exoskeletons with medical and military purposes. 

A first attempt to classify the Robots can be carried out for instance according to their application, kinematic chain and 
anatomy. In fact, the motion dynamics of a robot is described by means of more complex calculations \cite{ref5} and a more
specific approach would flee the scope of this work. However, physical concepts such as torque, linear and angular momentum,
acceleration, force and speed can be investigated during the construction of prototypes. From this point of view and with this 
motivation we showed the possibility of using robotics as a physics teaching tool, leaving to the educator the task of making
the simplest description and acessible to students whenever possible. 

From the educational point of view, the practical use of robotics is encouraging and some of the main reasons for its use 
as a teaching tool are: 
 
\begin{itemize}
\item[i)] It is an interdisciplinary science;

\item[ii)] Develops logical thinking, entrepreneurship, leadership, creativity and psychomotor ability;

\item[iii)] Provides teaching technologies related to sustainability, which is part of the new worldwide trends and Science 
Teaching;

\item[iv)] Allows students to apply the theory learned in classrooms to prove the importance of science in modern society 
and in the manufacture of products that makes life as we know it.

\end{itemize}

In addition, the use of simulations and computational modeling, which helps in the development of programming logic and in 
the subsequent process of architecture of the robotics project, facilitates the building up of relationships and meanings, 
promote constructivist learning \cite{ref6,ref7,ref8} and may also:

\begin{itemize}
\item[v)] Raise the level of cognitive process, requiring students to think at a higher level and generalizing concepts 
and relationships;

\item[vi)] Require students to refine their ideas more precisely;

\item[vii)] Provide opportunities for students to test their own cognitive models, detecting and correcting inconsistencies.

\end{itemize}

Once the robotics is considered an interdisciplinary area, it was possible to investigate physics concepts involved in the 
construction of the projects developed in this article. Thus, we use the fascination that most young people have for robotics
to encourage them to develop logical reasoning, problem solving and seek the understanding of various physical phenomena.

\section{APPLIED METHODOLOGY}

The PIBID has as main objectives the development of social and political teaching responsibilities, to provide the necessary 
assistance to facilitate the link between theory and practice, making the research a basic principle in education, and 
also stimulating the use of new information and communication technologies in the teaching-learning process. Thus, the main 
actions of the program are:

\begin{itemize}
\item[1)] Physical demonstrations of experiments with simple low cost materials, providing students the proof of the theory 
discussed in class;
\item[2)] The organization of games, pranks and thematic competitions based on topics of physics that increase students' 
curiosity and allow them to learn easily through plays;
\item[3)] The responsibility of choosing themes to be developed by means of workshops, drama, skits, exhibition and seminars 
or short lectures.
\end{itemize}

From this perspective we conducted a exhibition of Educational Robotics, which consisted in developing  programming logic 
and architecture systems similar to the biomechanics of living beings and also building combat robots, crawlers robots and 
automata systems for industrial use and probing environments. All this using only electronic scraps for the development of 
low-cost robots. Furthermore, during the exhibition, the following activities were also promoted: demonstration of robotics 
projects and prototypes, presentation of banners and videos, workshops and competition of combat robots.

Because a Educational Robotics Kit may have a high acquisition price, many schools in Brazil do not have the means to 
provide the necessary resources for students of the elementary and the high schools to acquire the materials to develop the 
activities we mentioned above. However, the use of electronic scraps \cite{ref9,ref10} offer a good alternative to build 
some prototypes.

The main focus of the activities reported in this paper was to relate the Robotics with Physics. Students who developed the 
prototypes had the opportunity to work on topics of physics such as electricity and electronics, with emphasis on building 
electrical circuits. The concepts of kinematics and dynamics were studied through the movement and interaction of the robot 
with the environment. The modern physics, for instance, could be worked out from the photoelectric effect that governs the 
operation of the light sensors (phototransistors), which also provide a way for describing the propagation of electromagnetic 
waves. The ultrasonic sensors of the robots, on the other hand, allowed the students to learn about the dispersion of sound 
waves.

Due to the large amount of built prototypes, it becomes unviable a detailed description of each project. However, we describe 
briefly some robots that were developed and we provide to reader to the references, which are freely accessible, that guided us 
in the construction and development of the projects discussed here.

\begin{figure}[!htb]
\centering
\includegraphics[scale=1,width=4cm]{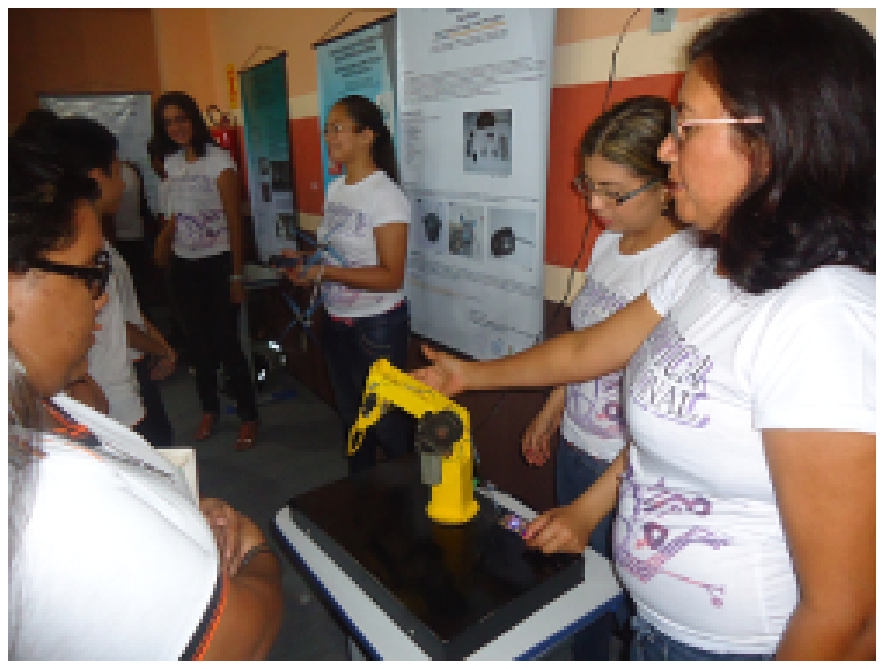}
\includegraphics[scale=1,width=4cm]{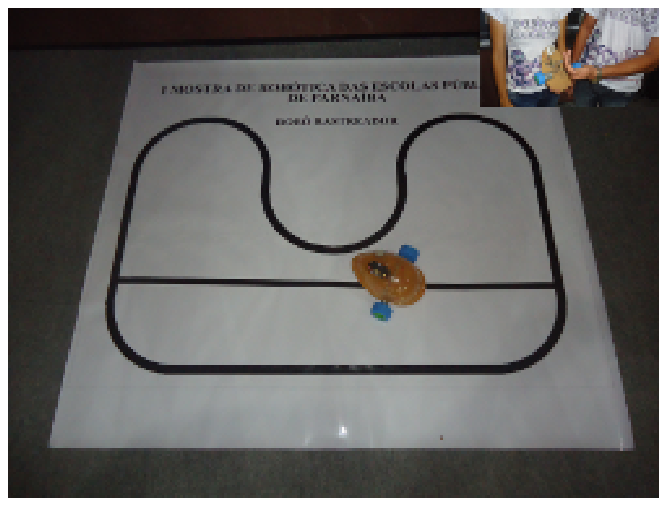}
\caption{The left figure shows students demonstrating how to operate a robotic arm while the right one shows a Follower Robot.} 
\label{figure1}
\end{figure}


In Fig.\ref{figure1} we have as an example the display of a robotic arm \cite{ref11}, built of wood, whose joints are driven 
by motors used in automotive power window. The central control it sits at the base and allows the arm to rotate 
$360\,^{\circ}$, move vertical and horizontal, open and close the jaws and the latter work by a DC motor 6V. Through this 
prototype was possible to explain concepts such as force, torque and levers. The follower robot \cite{ref12} follows the track 
highlighted by black using fototransitores positioned at the bottom of the chassis. It is constructed with plastic wheels 
supporting the chassis made of printed circuit board and is coated with a transparent mold for eggs Easter. Both projects do not
require programming.

On the other hand, by means of the intelligent control and free educational robotics systems, the use of Arduino plates 
also presents itself as a cheap alternative to build efficient robots with simplified programming that is easy to teach, 
and that have also a practical purpose. The Arduino is a platform for electronic prototyping, architected with a 
micro-controller Atmel AVR single board, embedded support for input/output, standard programming language C/C++, and 
a board that can transmit or receive data through a channel using a computer or other electronical devices \cite{ref13,ref14}.

\begin{figure}[!htb]
\centering
\includegraphics[scale=1,width=4cm]{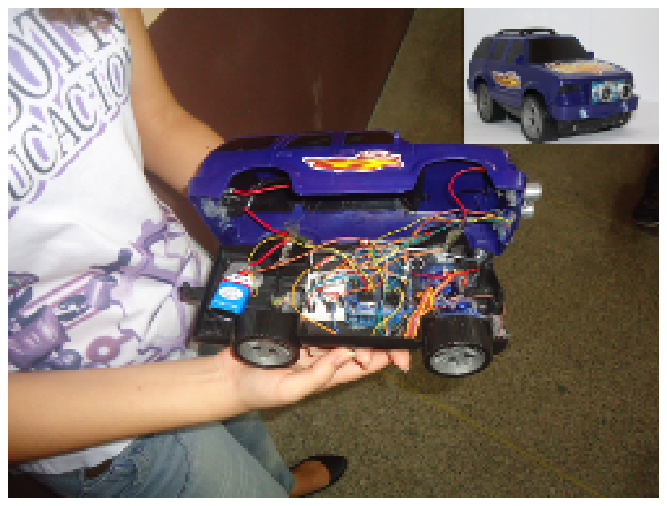}
\includegraphics[scale=1,width=4cm]{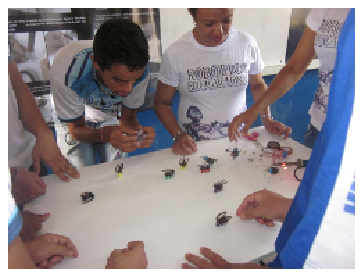}
\caption{The left figure shows a robot car with ultrasonic sensor and Arduino micro-controller while the right one shows 
an assembly plant where insect robots are built.} 
\label{figure2}
\end{figure}

The Fig.\ref{figure2} illustrates the two extremes: we have a robot car \cite{ref15} with ultrasonic sensor and Arduino 
microcontroller where the function of the sensor is to prevent the car collide with obstacles, works like sonar, which sends 
and receives sound waves to detect objects. The insect robot \cite{ref16,ref17} assembly workshop allowed students to interact 
with the activity. Being guided by PIBID fellows they could build, right there, a small insect endowed with incessant movement
using vibrate engine phones, toothbrush, batteries and wires. To improve the presentation of the project students have added 
wings to robots insects. 

\begin{figure}[!htb]
\centering
\includegraphics[scale=1,width=3.5cm]{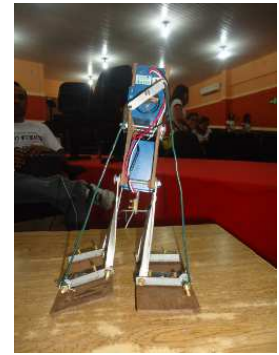}
\caption{This figure shows a biped robot.} 
\label{figure3}
\end{figure}

In Fig.\ref{figure3} we have the functioning of the biped robot \cite{ref18}. It was built using two servo motors as 
driving source, controlled by a satellite dish receiver and mounted on a wooden frame. The legs are made of aluminum bars 
fixed by screws to the feet which are also made of wood. Using Newton's laws and frictional force concepts, PIBID fellows 
might explain the robot motion that could walk by various surfaces and could make reclining lateral movements 
simulating a dance.

\begin{figure}[!htb]
\centering
\includegraphics[scale=1,width=4cm]{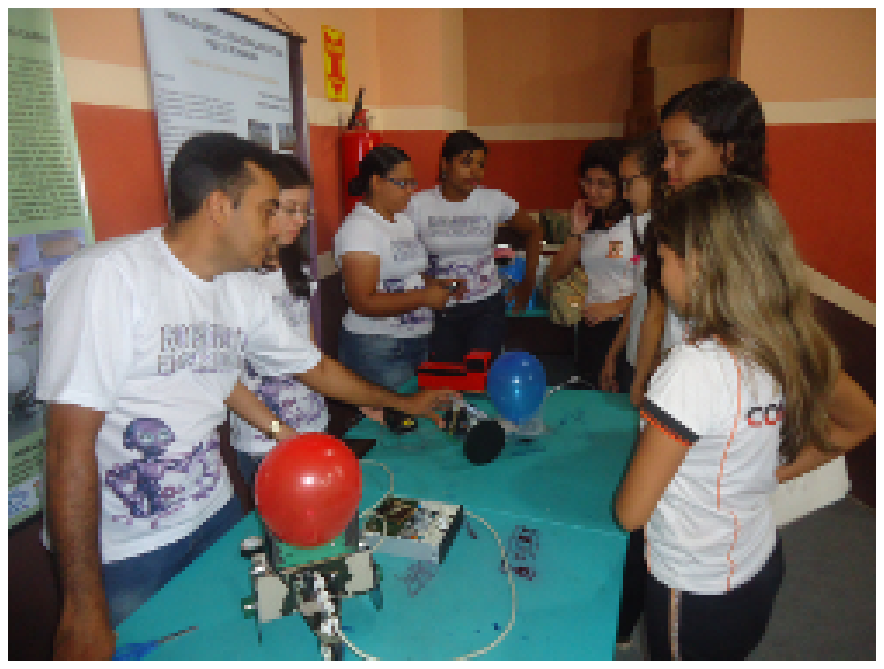}
\includegraphics[scale=1,width=4cm]{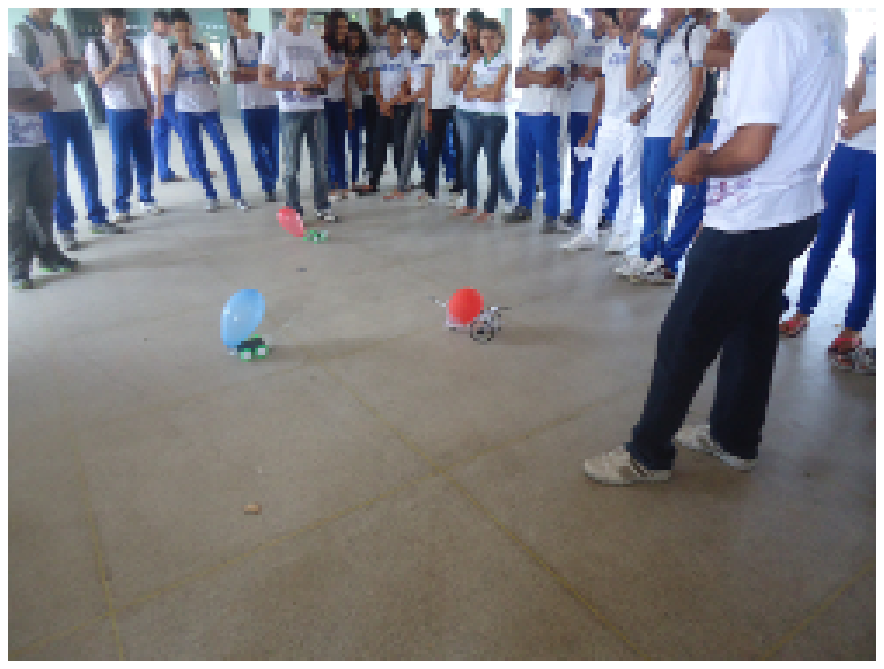}
\caption{The left figure shows students explaining the construction of fighting robots while the right one shows a competition 
between different robot prototypes.} 
\label{figure4}
\end{figure}

The Fig.\ref{figure4} shows the students of the schools field interacting with the fellows PIBID by combat robots 
\cite{ref19} competition. Robots have a wooden frame or acrylic, with plastic wheels made with CD or the movement thereof 
were made by 6V DC motors connected to a control connected to circuits mounted on the chassis. At times like these, 
for example, Physics topics such as dynamics and electricity were investigated.

In Table 1 we can find a list of the prototypes and Robotics projects developed by undergraduates students and that were 
exposed and used to teach topics of physics to the high school students. Some of the projects were taken and adapted from 
internet websites and articles while others from manuals. The Educational Robotics exhibition lasted one day in each school 
chosen to host the activities.

\begin{table}[h!]
\centering
\caption{List of prototypes and Robotics projects associated with the respective subjects of Physics worked.}
\begin{ruledtabular}
\begin{tabular}{p{2,5cm} p{5cm}}
ROBOT & THEME OF PHYSICS \\
\hline	
Robotic arm & Electrical circuits and Dynamics \\
Tracer robot & Modern Physics, Kinematics and Electromagnetic waves \\
Mouse robot & Electrical circuits and Kinematics \\
Biped robot & Electrical circuits and Dynamics \\
Cambate robot & Electrical circuits, Kinematics and Dynamics \\
Insect robot & Electrical circuits \\
Car robot & Electrical circuits, Undulating and Acoustics \\
Quadricopter & Electrical circuits, Dynamics and Hydrodynamics \\
Vehicle solar powered eco & Modern Physics, Electromagnetic waves, Electric Circuits, Kinematics and Dynamics \\
\end{tabular}
\end{ruledtabular}
\label{robotics}
\end{table}

\section{ACHIEVEMENTS}

The results reached through the activities are substancial and quite significant for the local reality of Parna\'iba as well 
as for the process of teacher training of undergraduate students and the development of their research skills in the field of 
education. The Educational Robotics exhibition was an alternative and innovative way to rise the interest of high school 
students by means of the field of science and technology. 

Furthermore, the PIBID students had the opportunity to create their teaching strategies along with programming logic, learning 
how to develop both biomechanics systems similar to the ones of living beings and robots of industrial utility and probing 
environments (see Table 1 for more). Remarkably, all these are low-cost robots built by the PIBID students using electronic 
scraps. These activities stablished a stronger link between the knowledge experienced in classrooms by undergraduate and high 
school students and their social environment, making clear the possible differences of the reality of these students.

It is worth stressing that the development and modeling of Robots along with the demonstration of physical principles using 
cheap materials of easy acquisition and that was promoted by PIBID through scientific shows, workshops, among others, provided 
students the proof of the theory discussed in class, contributing to their cognitive development and teaching practice.

Another important aspect to be mentioned was to encourage the use of programming languages, as these are tools increasingly 
used in several areas of current research, such as nuclear physics, condensed matter, particle and fields, etc. For this 
reason, the contact and use of these tools also contributed to the motivation of the undergraduate students 
who intend to pursue these areas in the future. 

The school teachers that also participated in the activities reported that their students began to interact more in their 
classes, frequently questioning about the role of science and its importance in society. Since recyclable materials were 
used in some prototypes, the activities was a way to call the community's attention to the advantages of recycling, also 
reinforcing the importance of sustainability through teaching.

\section{CONCLUSION}

The most modern aspects of education advocate a focused education to build a dynamic mentality and constructivist by 
the students. Based on these aspects, this paper explored the possibility of using interdisciplinary themes, in relation 
to the teaching of physics, be treated with an innovative methodology by building and exposure robotics projects. This 
new methodology brought a big challenge for students since they had to learn facing the poor condition (or the lack of it) 
of the science lab in their schools. For instance, not all prototypes had micro-controller, most were architected so that 
we had a dynamic and autonomous physical structure driven by electrical and electronic circuits. 

Despite of this, many difficulties were overcome and we expect that the implementation of a programming logic for all projects 
and the development of more sophisticated chassis and structures can come into reality in future activities. 

Our hope is that the great results presented here encourage other groups, supervised by university teachers, to establish 
similar activities in order for science to be disseminated among the young students. By doing this, we are able to introduce 
the technological reality of the current society to those students, making them feel included and stimulated to become future 
teachers and scientists.

\begin{acknowledgments}

The authors thank CAPES for financial support. Our special thanks to all fellows PIBID, teacher Adriana Rocha for 
collaboration, teachers Ademar Ribeiro, Josenildo Silva and Orlando Diniz for support to students in high schools 
and Herondy Mota of Physics and Astronomy Department from University of Sussex by the reviewing the text. 

\end{acknowledgments}


\end{document}